\documentclass[aps,pre,twocolumn,superscriptaddress,showkeys,showpacs,longbibliography]{revtex4-2}
\usepackage[english]{babel}
\usepackage{amsmath}
\usepackage{amssymb}
\usepackage{comment} 
\usepackage[version=4]{mhchem}
\usepackage{etoolbox}
\usepackage{mwe}
\usepackage{graphicx}
\usepackage{dcolumn}
\usepackage{xcolor}
\usepackage{bm}
\usepackage{float}

\newcommand{\dg}{^{\dagger}}
\newcommand{\bk}{{\bf k}}
\newcommand{\bx}{{\bf x}}
\newcommand{\pmat}[1]{\begin{pmatrix} #1 \end{pmatrix}}
\usepackage{hyperref}
\usepackage[capitalize]{cleveref}
\hypersetup{
    colorlinks=true,
    linkcolor=blue,
    filecolor=blue,      
    urlcolor=blue,
    citecolor=blue,
    pdftitle={draft2},
    pdfpagemode=FullScreen,
    }
\usepackage[all]{hypcap}    
\usepackage[capitalize]{cleveref}

\begin{document}

\preprint{APS/123-QED}

\title{A higher dimensional generalization of the Kitaev spin liquid }

\author{Po-Jui Chen}
\email{pc863@physics.rutgers.edu}
 \affiliation{%
Department of Physics and Astronomy, Rutgers University, Piscataway, New Jersey 08854, USA 
}%
\author{Piers Coleman}%
\affiliation{%
 Department of Physics and Astronomy, Rutgers University, Piscataway, New Jersey 08854, USA
}%
\affiliation{Department of Physics, Royal Holloway University of London, Egham, Surrey TW20 0EX, United Kingdom}

\date{\today}

\begin{abstract}

We construct an exactly solvable model of a four-dimensional Kitaev spin liquid. The lattice structure is orthorhombic and each unit-cell contains six sublattice degrees of freedom. We demonstrate that the Fermi surface of the model is made up of two-dimensional surfaces. Additionally, we evaluate the energy cost of creating visons using scattering theory. The positive bond-flip energy suggests that the flux-free state is locally stable. Our model sheds light on the realization of high-dimensional fractionalization.



\end{abstract}

\maketitle


\section{Introduction}

The quantum spin liquid is an exotic phase of matter 
 which lacks long-range magnetic order and does not exhibit any broken symmetry. Instead, the ground state of a spin liquid is characterized by its long-range quantum entanglement\cite{Anderson1987,Anderson1973,Savary2016,balents_spin_2010,Wen2019}. Another useful way to understand quantum spin liquids is in terms of their elementary excitations. The low-energy effective field theory for quantum spin liquids often involves fractionalized spin excitations moving under an emergent gauge field. Here, fractionalization refers to the phenomenon whereby elementary excitations carry fractional quantum numbers: typical examples include neutral spin 1/2 spinons, and spinless visons, which describe excitations of the gauge fields. Although quantum spin liquids provide a new venue for physicists to explore quantum materials and solve fundamental problems in strong correlation physics, it is a challenge to discover exactly solvable models that can be used as platforms for studying quantum spin liquids. 
\begin{figure*}[t!]
   \includegraphics[width = 0.9\linewidth]{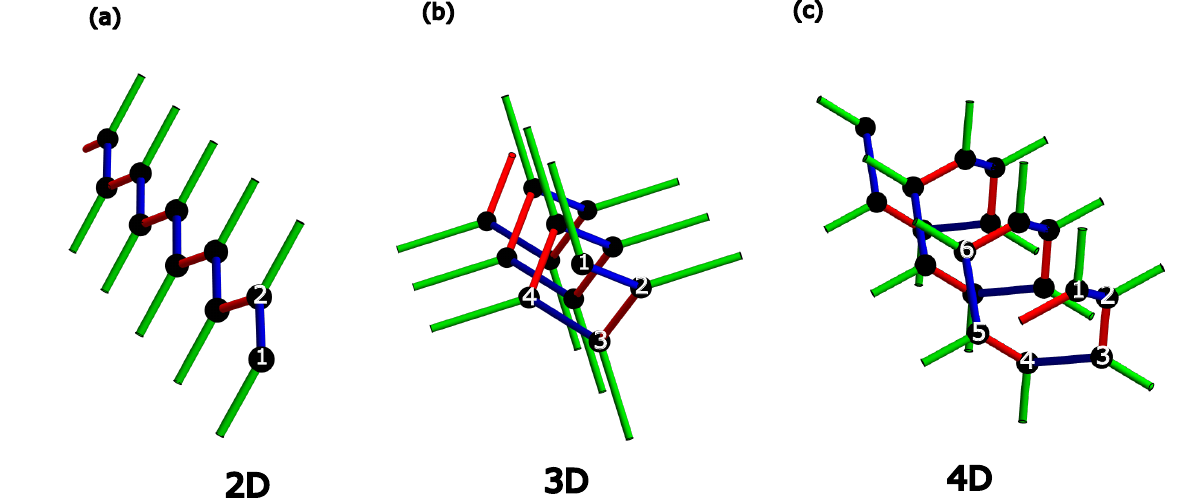}
    \caption{The dimensional evolution of spiral Kitaev models.  (a) the honeycomb represented as a 2D flattened spiral (b)  the 3D hyper-octagon and (c) the 4D hyper-hexagon, illustrated in three dimensions by projecting out the $\hat n = (1,1,1,0) $ direction. The blue-red bonds denote the alternating $xx,yy$  chain, while the green $zz$ bonds link the spirals.  }
    \label{dim}
\end{figure*}
One of the great innovations in this respect is  Kitaev's  exactly solvable honeycomb model \cite{KITAEV20062} , which consists of bond-dependent Ising interactions between spin-1/2 local moments. Kitaev's decomposition reveals the underlying physics of spins, fractionalized into Majorana fermions moving through a static $Z_2$ gauge field. The defects of the $Z_2$  gauge field are plaquette fluxes, called visons. Kitaev's model demonstrated that the transition between gapless
and gapped topological phases is driven by tuning the strengths of the spin coupling constants. The gapped phases of the Kitaev honeycomb model support the existence of non-abelian anyons of possible interest to topological quantum computation\cite{Lahtinen2017,Nayak2008}. Since the original work by Kitaev, many exactly solvable three-dimensional variants of the original model have been studied\cite{M2009,Hermanns2014,Hermanns2015,O'Brien2016,Yamada2017,Eschmann2020,Jahromi2021}, classified through their topology and the presence of gapless Dirac cones, or Fermi surfaces in the excitation spectrum\cite{O'Brien2016,Yamada2021}. There are also several theoretical proposals which extend the Kitaev model by introducing an additional orbital degree of freedom\cite{Coleman2022,Tsvelik2022,Panigrahi2024}. 

On the experimental side, several experiments have sought the realization of the Kitaev spin liquid in materials with strong spin-orbit coupling\cite{Do2017,PhysRevB.95.144406,Winter2017,Hermanns2015}.  $\alpha$-RuCl$_3$ and iridates \cite{Do2017,PhysRevLett.114.077202} are candidates for Kitaev materials, in which the interplay between strong correlation in $d$ electrons and spin-orbit coupling gives rise to the anisotropic Ising interaction. In these Kitaev materials, inelastic neutron scattering\cite{Banerjee2017} and Raman scattering \cite{Glamazda2016,S2015,Panigrahi2024}  have revealed tentative signatures of spin fractionalization in these materials.

On the other hand, there has been a revival of interest in high-dimensional physics. In particular, the four-dimensional quantum hall state\cite{doi:10.1126/science.294.5543.823}, topological Anderson insulator\cite{PhysRevB.108.085306}, Floquet topological insulator\cite{PhysRevB.109.125303}, and five-dimensional Weyl semimetals\cite{PhysRevB.95.235106,PhysRevB.94.041105} have been subjects of theoretical investigation. Experimentally, atomic systems\cite{PhysRevLett.115.195303}, photonic platforms\cite{PhysRevA.87.013814,PhysRevA.93.043827}, and specially designed quantum  circuits \cite{wang_circuit_2020,yu_4d_2020-1} have been proposed to simulate 4D physics. These experimental advances provide an additional    motivation for studies of higher dimensional physics.  
    
     In an effort to gain further insight into Kitaev spin liquids and high-dimensional physics, here we introduce a four-dimensional realization, a natural generalization of the 2D and 3D Kitaev spin liquids, as shown in Fig. \ref{dim}(a) and (b),  which we call the hyper-hexagonal lattice.  The key structural motif in our  model is  a chain of alternating xx and yy bonds, which we then wrap around an even-sided helix, to form a structure with an even number of $2n$ atoms per unit cell, as shown in Fig. \ref{dim}.  We then add zz-bonds as cross-links between the helices.  Thus, the 2D honeycomb model consists of zig-zag chains, connected by zz cross-links.  For $n=2$, we create a four-sided helix with four atoms per unit cell, forming the three-dimensional ``hyperoctagonal'' lattice\cite{Hermanns2014}.  Generalizing this procedure, for a $2n$-sided helix  we form a structure with cross-links in $n$ independent directions, which together with the axis of the helix forms a $D=n+1$ dimensional structure. Thus for $n=3$, we form a  six-sided helix with a hexagonal cross-section embedded in $D=4$ dimensions - the ``hyper hexagon".   To visualize this novel structure, we project it into the 3-dimensional hyperplane perpendicular to the $\hat n=(1,1,1,0)$ direction: viewed along the axis of the helix, it then forms a hexagonal structure in which the xx-yy chain wraps around the hexagon as one proceeds along the fourth w-axis,  (0,0,0,1) as shown in Fig. \ref{dim}(c).       
    
  With this lattice geometry, we can perform a spin-fermionization in four dimensions, formally identical to those in two and three dimensions. The resulting physics describes a free Majorana fermion moving on a four-dimensional lattice containing six sites per unit cell. Its exact solvability enables us to investigate its energy spectrum, ground state, and elementary excitations.

The remaining discussion is organized as follows. Section \ref{sec2} describes the lattice geometry of the four-dimensional Kitaev spin liquid, including its  Hamiltonian, and Wilson loop operators. In Section \ref{sec3} , we compute the energy spectrum for the isotropic case and demonstrate that the Fermi surface of excitations is two-dimensional, embedded in a four-dimensional manifold. In Section \ref{sec4} , we discuss the vison gap of our model and argue that the system is locally stable against gauge excitations. Section \ref{sec5} discusses the validity of the Jordan-Wigner transformation for the four-dimensional model, the possibility of arbitrary dimensional extensions.




\section{The model and its spin-fermionization}\label{sec2}

The Kitaev spin model
\begin{equation}\label{eq1}
    \begin{split}
        H_{K} = \sum_{\langle i,j\rangle} J_{\alpha_{ij}} \sigma^{\alpha_{ij}}_{i}
       \sigma^{\alpha_{ij}}_{j},
    \end{split}
\end{equation}
where  $\langle ij\rangle$  denotes neighboring sites on the lattice and  ${\alpha_{ij}}$ identifies the type of bond ($xx,yy,zz$), can be embedded on any lattice with three-fold co-ordination that can be consistently tiled with $x$, $y$ or $z$ bonds. Its extension to 4 dimensions requires that we now construct such a lattice.

\begin{figure}[h]
   \includegraphics[width =8cm]{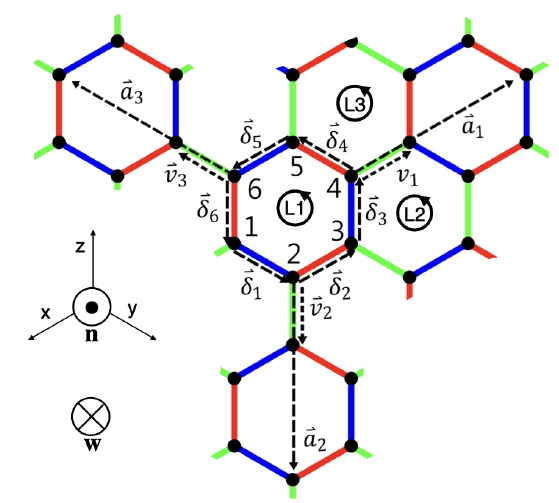}
   \caption{(a)The projected lattice structure of our 4D Kitaev spin liquid. The figure is obtained by first projecting along the $w$ direction and then projecting along the ${\bf n}= (1,1,1,0)$ direction. (b) Three fundamental {plaquette} operators for the 4D Kitaev spin liquid. There are 12 or 14 bonds in a fundamental loop. Emitter and Absorber specify the sign arrangement of a loop for the flux-free configuration.} \label{Fig2}
\end{figure}

We begin by introducing the detailed helical spatial structure for our four-dimensional spin liquid, which we call the hyper-hexagonal lattice. To this end, we consider an orthorhombic lattice with six lattice points in one unit cell. as  shown in Fig.\ref{Fig2} .  We take our four dimensional basis to be
${\bf \hat x} = (1,0,0,0)$, ${\bf \hat y} = (0,1,0,0)$,  ${\bf \hat z} = (0,0,1,0)$ and  ${\bf \hat w} = (0,0,0,1)$.
To display the structure, we have projected it into the two-dimensional plane perpendicular to the $\hat {\bf w} $ and ${\bf n} = (1,1,1,0)$ directions. The projected  $\bf \hat x$, $ \bf \hat y$ and $\bf \hat z$ directions are shown as a right-angled coordinate system, such that  ${\bf \hat n}$ is perpendicular to the plane of the paper. We label the six sites in the unit cell by $i= 1, 2, \dots 6$. Each site is connected to its nearest neighbors at  $i' = (i\pm 1 ){\rm mod} 6$. On the other hand, each unit cell is connected to its six neighbors as shown in Fig.\ref{Fig2}, and labeled by green bonds, which enables the lattice structure to expand in the $x,y,z$, linking direction. From  Fig. \ref{Fig2}, site  $i$ links to site $(i+3) {\rm mod}(6)$ in neighboring unit cells, i.e 1 is connected to 3, 2 to 5, and 3 to 6 by the green inter-cell links. 

The geometric structure within a unit cell is described by vectors 
$\boldsymbol{\delta_i}= {\boldsymbol x}_{i+1}-{\boldsymbol x}_{i}$, ($i=1,6$), where $\bm{x_i}$ denotes the position of atoms sitting inside the unit cell and \begin{eqnarray}
\boldsymbol{\delta_1} = \hat{\bf y}+{\bf w} &, \ & \boldsymbol{\delta_4}=-\hat{\bf y}\cr
\boldsymbol{\delta_3} = \hat{\bf z}+{\bf w} &,& \boldsymbol{\delta_6}=-\hat{\bf z}\cr
\boldsymbol{\delta_5} = \hat{\bf x}+{\bf w} &,& \boldsymbol{\delta_2}=-\hat{\bf x}.
\end{eqnarray}
With this choice of lattice vectors, the pitch of the spiral is 3 units in the $w$ direction.
The vectors representing the green bonds linking the unit cells are
\newcommand{\bn}{\bf n}
\newcommand{\by}{\bf y}
\newcommand{\bz}{\bf z}
\begin{equation}\label{green}
    \begin{split}
        \boldsymbol{v_1} &= \frac{1}{2}{\bn}  - \hat \bx,\\
         \boldsymbol{v_2} &= \frac{1}{2}\bn - \hat \bz,\\
          \boldsymbol{v_3} &= \frac{1}{2}\bn - \hat \by \\
    \end{split}
\end{equation}

Notice how even-numbered bonds $\bm{\delta _i}$, ($i=2,4,6$) and cross-links $\boldsymbol{v_i}$ ($i=1,3$) do not advance in the $w$ direction.  This choice has been chosen to guarantee that passage around any hexagonal loop in the structure increments the $w$ co-ordinate by a multiple of 3, leading to a closed structure. 

The positions of unit cells are described by the following primitive lattice vectors
\begin{equation}\label{bravis}
    \begin{split}
        \boldsymbol{a_1} &= (\frac{-3}{2},\frac{3}{2},\frac{3}{2},2),\\ 
\boldsymbol{a_2} &= (\frac{3}{2},\frac{3}{2},\frac{-3}{2},-1),\\ 
\boldsymbol{a_3} &= (\frac{3}{2},\frac{-3}{2},\frac{3}{2},2),\\ 
\boldsymbol{a_4} &= (0,0,0,3),\\ 
    \end{split}
\end{equation}
Note that ${\bf a}_1+{\bf a}_2+{\bf a}_3- {\bf a}_4 = \frac{3}{2} \bn$, so we could have also chosen $\frac{3}{2}\bn$ as a lattice vector. 

Consistency requires that after passage around a closed loop in the projected 2D space, as shown in Fig.\ref{Fig2},  the final coordinate $\bm{x_f}$ either coincides with the initial point $\bm{x_i}$ or a point that is shifted by a Bravais lattice vector, i.e $\bm{x_f}-\bm{x_i}=\sum_j n_j\bm{a_j}$, with $n_j\in \mathbb{Z}$. In particular, we note that advancing anticlockwise around the primary hexagonal loop (L1 in Fig.\ref{Fig2}), gives a pitch of $\circlearrowleft_1=\sum_{i} {\boldsymbol{\delta}_i}= {\bf a}_4$ while advancing around the secondary loop L2 gives $\sum_{\circlearrowleft_2} = -(\bm{\delta_1}+\bm{\delta_3}+\bm{\delta_5}+\bm{v_1}+\bm{v_2}+\bm{v_3})= -\frac{3}{2} \bn - {\bf a}_4$ and $\sum_{\circlearrowleft_3} =\bm{v_1}+\bm{v_2}+\bm{v_3}-(\bm{\delta_2}+\bm{\delta_4}+\bm{\delta_6})=\frac{3}{2}
 \bn $.  

Having identified the Bravais lattice vectors, we can determine the  volume of the unit cell
\begin{equation}
     \Omega_{4D} =  \epsilon^{\alpha\beta\gamma\delta}
{a_1^{\alpha}}
{a_2^{\beta}}
{a_3^{\gamma}}
{a_4}^{\delta}=\frac{81}{2},
\end{equation}
and the corresponding 
corresponding reciprocal lattice vectors, normalized so that ${\bf b}_a\cdot {\bf a}_b =  \delta_{ab} $, determined from 
${\bf b_1}^{\alpha} = \frac{1}{\Omega_{4D}}\epsilon^{\alpha\beta\gamma\delta}
     {a_2^{\beta}}
{a_3^{\gamma}}
{a_4}^{\delta}$ and cyclic permutations, which gives 
\begin{equation}\label{bvec0}
    \begin{split}
        {\bf b_1} &=(0,\frac{1}{3},\frac{1}{3},0)\\
        {\bf b_2} &= (\frac{1}{3},\frac{1}{3},0,0),\\
        {\bf b_3} &= (\frac{1}{3},0,\frac{1}{3},0),\\
       {\bf b_4} &= (-\frac{1}{9},-\frac{1}{9},-\frac{4}{9},\frac{1}{3}).\\
    \end{split}
\end{equation}
It is convenient to resolve the momentum in reciprocal space ${\bf k} = \sum _{\alpha = 1,4} k_{\alpha } {\rm b}_{\alpha} $, where the components are then given by $k_{\alpha} = {\bf k}\cdot {\rm a}_{\alpha}$.

\subsection{Majorana representation}

Here, we briefly review the application of Kitaev's method to the hyper-hexagon. The central idea of Kitaev's method is to fractionalize spins into Majorana fermions, according to  $ \sigma^{\alpha}_{i} = 2ic_j b^{\alpha}_j$ , where $c_j,b_j^{\alpha}$ are Majorana operators whose particles are their own anti-particle $c_j = c_j^{\dagger},b^\alpha_j = b_j^{\alpha\dagger} $. These operators also obey the commutation relations $\{c_i,c_j\} = \delta_{ij},\{b^\alpha_{i},b^\beta_{j}\} = \delta_{\alpha,\beta}\delta_{i,j},\{c_i,b^\alpha_{j}\} = 0$. One should notice that when spin operators are replaced as a product of Majorana operators, the local Hilbert space of the system is expanded, necessitating a projection into the physical subspace through the constraint $D_j =  4ic_jb^x_jb^y_jb^z_j=1$. After this substitution, the resulting Hamiltonian becomes
\begin{equation}\label{theham}
    H_{K} = \sum_{}J_{\alpha_{ij}}u_{ij}(2ic_{i}c_{j}),
\end{equation}
where $u_{ij} = -2ib^{\alpha_{ij}}_ib^{\alpha_{ij}}_j $  satisfies the unitary condition $u_{ij}^2 = 1$. If we exchange the index $i,j$, the operator picks up a minus sign $u_{ij} = -u_{ji}$. This factor characterizes the emergent $Z_2$ gauge symmetry under the transformation
 $c_i\rightarrow Z_i c_i, u_{ij}\rightarrow  Z_iu_{ij}Z_j$.

 \begin{figure}[h]
   \includegraphics[width =8cm]{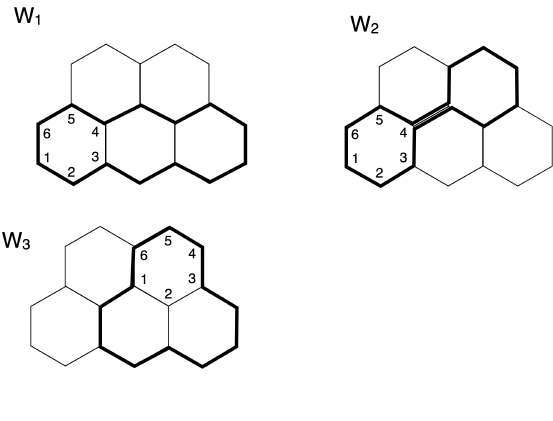}
   \caption{ Three  fundamental Wilson loops for the 4D Kitaev spin liquid. There are 12 or 14 bonds in a fundamental loop. Note that $W_1 W_2 = W_3$} \label{Fig3}
\end{figure}
Another important aspect of this model is its conserved quantities. We can find the fundamental Wilson loop operators that commute with the Hamiltonian. In the hyper-hexagon,  these loops are 12 or 14 lattice constants long, as shown in Fig.\ref{Fig3} .The loop operators are  products of Pauli matrices along the selected path. 
 \begin{equation}
     \begin{split}
        W_p = \prod_{i\in \text{loop } p}\sigma^{\alpha_i}_i,
     \end{split}
 \end{equation}
 where $i$ denotes the site index while the component $\alpha_i\in[x,y,z]$ is chosen from the  bond external to the loop at site $i$ (so $\alpha_i \neq \alpha_{i,i+1}, \alpha_{i-1,i}$).  
As in the hyper-octagon,  these loops are not independent of each other\cite{Hermanns2014}, satisfying $W_3 = W_1W_2$.  
 Both operators are constants of motion since  $[W_p,H]=0$. In terms of the Majorana language, they can be rewritten as $W_{p} = \prod_{<ij>\in loop}u_{ij}$. The loop operators square to unity, with eigenvalues $\pm 1$, corresponding  to a $Z_2$ flux. In what follows, we provide the guiding rule of gauge choice on bonds $u_{ij}$ such that the system is flux-free. Our construction relies on the definition of emitter and absorber, corresponding to odd and even numbered sites, respectively.  On the fundamental loops, we specify a flux-free configuration(all $W_p$ s are 1) by assigning an alternating emitter-absorber arrangement. With this pattern, we define the ordered bond variables as follows
\begin{equation}
        u_{\langle ij\rangle} = 
        \left \{
        \begin{aligned}
              u_{ij},\text{if $i$ is odd}\\
             u_{ji},\text{if $j$ is odd}\\
        \end{aligned}
        \right.
\end{equation}
With this definition, we can express the loop operators as $\prod_{\langle ij\rangle\in loop}u_{\langle ij\rangle}$ the gauge choice for the flux-free sector is $u_{\langle ij\rangle}=1 $ for all $\langle ij\rangle$ pairs. \\

\section{Majorana band structure}\label{sec3}

\subsection{Energy spectrum }
 We are now in a position to rewrite the gauge-fixed Hamiltonian in momentum space. We first transform  the Majorana fermions into momentum space operators  $c_{\alpha,\bk} = \frac{1}{\sqrt{N}}\sum_{j}e^{-i\bk\cdot \boldsymbol{R_j}}c_{\alpha,j}$,   where $\alpha=(1,2,\dots 6)$ denotes the sublattice index, $\bf k$ is the momentum and  ${\bf R}_i$ denotes the location of the unit cell.  The momentum space operators  $c_{\alpha,\bk}$ behave as canonical complex fermions, namely
 $\{c_{\alpha,\bk},c\dg_{\beta,\bk'\}= \delta_{\alpha\beta}\delta_{\bk,\bk'}} $, however, they are not
 independent in the two halves of momentum space, since $c\dg_{\alpha,\bk} = c_{\alpha,-\bk}$. 
  We can then write the Hamiltonian as \begin{equation}
  H = \sum_{\bk\in \frac{1}{2}{\rm BZ}} c\dg_{\bk} h({\textbf{k}})c_{\bk},\end{equation}
  where ${c}\dg_{\bf k} = (c\dg_{1 \bk}, c\dg_{2\bk}, c\dg_{3\bk}, c\dg_{4\bk}, c\dg_{5\bk}, c\dg_{6\bk})$ and the sum over momentum is restricted to one half the Brillouin zone.  Assuming a  flux-free configuration,  
\onecolumngrid

\begin{equation}
       h(\textbf{k}) =
        \pmat{
        0 & J_x i& 0 & J_z ie^{-i{k_1}}&0 & J_y ie^{-i {k_4}}\cr 
        -J_x i&0 &-J_y i&0&-J_zie^{i k_2}&0\cr
        0 & J_yi&0 & J_xi&0 & J_z i e^{-i{k_3}}\cr
        -J_zie^{i{k_1}} & 0&-J_x i&0&-J_y i&0\cr
        0 & J_zie^{-i{k_2}}&0 & J_y i&0 & J_x i \cr
        -J_y i e^{i {k_4}}&0& -J_z ie^{i  {k_2}}&0&-J_x i&0 },
\end{equation}
\noindent{where the factor of 2 in the terms $2i c_i c_j$ in \eqref{theham} has been absorbed into the antisymmetrization of the matrix. All momenta are resolved along the reciprocal lattice vectors, $\bk = \sum_{a=1,4}k_a\bm{b_a}$, where  $k_a = \bk\cdot  \boldsymbol{a_a}$.   As in lower dimensions, the 4D Kitaev model has a sublattice symmetry linking odd and even sites. If we redefine the spinor to separate the odd and even sectors, $\tilde{c}\dg_{\bf k} = (c\dg_{1 \bk}, c\dg_{3\bk}, c\dg_{5\bk}, c\dg_{2\bk}, c\dg_{4\bk}, c\dg_{6\bk})$, then $H = \sum_{\bf k\in \frac{1}{2}{\rm BZ}} \tilde{c}\dg_{\bf k} {\cal H}({\bf k}) \tilde c_{\bf k}$ acquires a compact, off-diagonal structure,  
\begin{equation}\label{HQ}
   {\cal H}(\bk) = \pmat{
     & i Q(\bk) \\
    -i Q\dg(\bk) & 
},\qquad
        Q(\bk) = \pmat{J_x & J_ze^{-i{k_1}}&J_y e^{-i {k_4}}\cr 
        J_y&J_x&J_ze^{-i{k_3}}\cr
        J_ze^{-i{k_2}}&J_y&J_x 
        }.
\end{equation} 
We note that the chiral operator $\Lambda= \sigma_z\otimes 1_3 $ anticommutes with $\Lambda$, $\{{\cal H}(\bk),\Lambda\}=0$, guaranteeing that the spectrum is particle-hole symmetric. As in the Kitaev honeycomb model, the absence of next-nearest hopping terms is connected with the time-reversal symmetry of the underlying spin model. When we introduce Zeeman coupling terms, the equivalent Majorana representation develops next-nearest neighbor hopping which break this symmetry.    

\twocolumngrid
We now discuss the spectrum for the case of isotropic coupling constants ($J_x=J_y=J_z$)  via the eigenvalue equation ${\rm det}[E-H(\bk)]={\rm det}[E^2-Q\dg(\bk)Q(\bk)]=0$. Fig. \ref{Fig4}(a)  shows the band structure along the closed path $\Gamma\rightarrow M \rightarrow X \rightarrow K \rightarrow L \rightarrow \Gamma$ , where  $\Gamma$ is the origin while $M = \frac{1}{2}\boldsymbol{b_1},
X =\pi (\boldsymbol{b_1} - \boldsymbol{b_2}),
K =\pi(\boldsymbol{b_1} - \boldsymbol{b_2}+ \boldsymbol{b_3})$ and $
L = \pi(\boldsymbol{b_1} - \boldsymbol{b_2}+ \boldsymbol{b_3}+ \boldsymbol{b_4})$. There is a fourfold degeneracy at the $\Gamma$, $M$ and $L$ points which each connect a flat band to two linearly dispersing excitations.
Along the high symmetry lines, the flat bands can be understood from the matrix structure of the Hamiltonian. For momenta located along the lines  $\overrightarrow{\Gamma M}$ or $\overrightarrow{L\Gamma}$, the Q-matrix is given by
\begin{equation}
    \begin{split}
        Q(\bk\in\overrightarrow{\Gamma M}) &= 
        J\left(
        \begin{array}{ccc}
            1 & e^{-  i s} &1\\
             1&1 &1\\
             1&1&1
        \end{array}
        \right),\\
        &\text{$\bk=s\bm{b_1},0\le s \le \pi $},\\
          Q(\bk\in\overrightarrow{\Gamma L}) &= 
        J\left(
        \begin{array}{ccc}
            1 & e^{  it} &e^{ i   t}\\
             1&1 &e^{ i   t}\\
             e^{-i  t}&1&1
        \end{array}
        \right),\\
        &\text{$\bk=-t(\bm{b_1}-\bm{b_2}+\bm{b_3}+\bm{b_4}),0\le t \le \pi$}.\\
    \end{split}
\end{equation}
and the corresponding determinants vanish along these lines.  In fact, these flat bands are part of a two-dimensional Fermi surface, as we now discuss. 

\begin{figure}[h]

   \includegraphics[width =7cm]{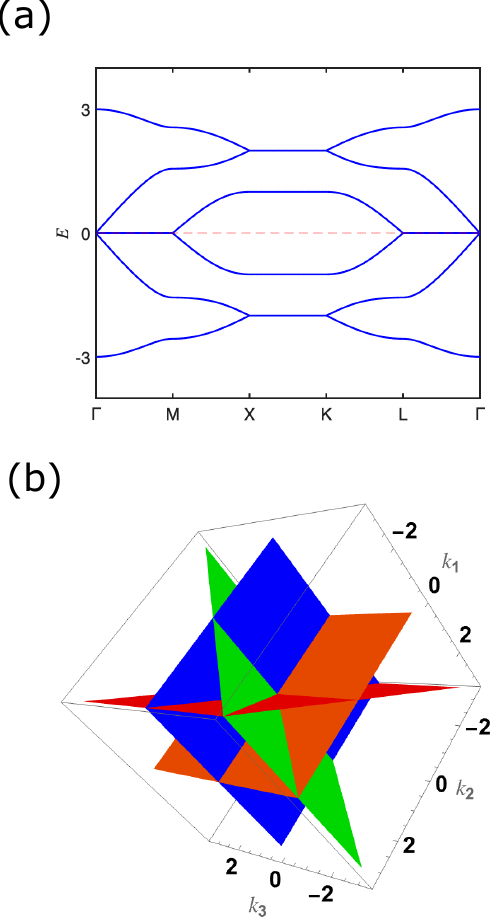}

    \caption{(a) The spectrum along a high symmetry momentum path $\Gamma\rightarrow M \rightarrow X \rightarrow K \rightarrow L \rightarrow \Gamma$. (b) Four of the six two-dimensional Fermi surfaces defined in \ref{sixfs}, projected into the first three dimensions $(k_1,k_2,k_3)$. }  \label{Fig4}
\end{figure}

\subsection{Fermi Surface}

 {At the Fermi energy $E=0$, the eigenvalue equation is $\det [Q]=0$, which implies two independent constraints,  $\text{Re}[\det[Q]]=0$ and $\text{Im}[\det[Q]]=0$, whose intersection defines a two-dimensional surface embedded within the four-dimensional Brillouin zone. As in lower dimensions, this surface of  zero modes is protected via the an underlying projective symmetry\cite{Yamada2017,O'Brien2016}} of the model.  If we observe a slice of  this  Fermi surface projected into three dimensions at fixed $k_4=t$, we obtain a string in three dimensions $\vec k(\theta,t)$, where $\theta$ defines the position along the string: as $t$ is varies,
 the string sweeps out a surface. 

{For the isotropic case, we can determine the Fermi surface by examining the internal matrix structure of $Q$. By identifying the points in momentum space where one row (or column) becomes linearly dependent on the others. In fact, for the isotropic case, we can obtain all possible solutions just by demanding that any two columns or two rows are proportional to one-another. This gives rise to {\sl six} separate surfaces  of the following form
\begin{equation}\label{sixfs}
\begin{split}
   (k_1,k_2,k_3,k_4) =  \left\{
        \begin{aligned}
         (t,s,0,t),\\
          (s, -t, t, t),\\
            (0,s,t,t),\\
             (t,-t, s, t), \\
             (0,0,s,t),\\
              (s,0,0,t),
        \end{aligned}
        \right.
\end{split}
\end{equation}
Fig. \ref{Fig4}(b) shows the first four of these  Fermi surfaces projected into three dimensions.  }

\subsection{Density of states}

{
In the absence of the Van-Hove singularity, a two-dimensional Fermi surface in a four-dimensional space leads to a density of states that is linear in energy $N(E)\propto E$.  
 The eigenvalue equation is  $\det[E^2-Q(\bk)^\dagger Q(\bk)]=0$ . For a generic  momentum on the Fermi surface where $E=0$, $\det[Q(\bm{k_0})]=0$, and the low-energy expansion in the vicinity of the Fermi surface at $\bk = \bk_0+\delta\bk $ is given by 
\begin{equation}
    E(\bk) = \pm \sqrt{(\delta\bk\cdot\nabla Q_1(\bk_0))^2+(\delta\bk\cdot\nabla Q_2(\bk_0))^2}.
\end{equation}
where   $Q(\bk)=Q_1(\bk)+iQ_2(\bk)$ separates the real and imaginary parts of the determinant. The density of states is then given by 
\begin{equation}
    D(E) = \int \frac{d^4k}{(2\pi)^4}\delta (|E| -\sqrt{(\delta\bk\cdot\nabla Q_1)^2+(\delta\bk\cdot\nabla Q_2)^2})
\end{equation}
If we resolve the momenta into components parallel and perpendicular to the Fermi surface, writing $d^4k = d^2 k_0 d^2 k^{\perp}$, decomposing $\bk^{\perp} =k^{\perp}_1 \hat {\bf e}_1 +k^{\perp}_2 \hat {\bf e}_2 $, where $\hat {\bf e}_{1,2}$ are orthogonal normals to the Fermi surface, then we can write 
\begin{eqnarray}
    D(E) &=& \int \frac{d^2 k_0}{(2\pi)^2}D(E,\bk_0)\cr
    D(E,\bk_0)&=& \int \frac{d^2 k^{\perp}}{(2\pi)^2}
    \delta (|E| -\sqrt{(\delta\bk\cdot\nabla Q_1)^2+(\delta\bk\cdot\nabla Q_2)^2})\cr&&
\end{eqnarray}
Changing variables to $u=\delta\bk\cdot\nabla Q_1(\bk_0) $ and $v=\delta\bk\cdot\nabla Q_2(\bk_0)$, then 
$
    d^2k^{\perp} = {dudv}/{{\rm det}[\hat e^a\cdot \vec\nabla Q_b]}.
$
Switching to polar co-ordinates $u+iv = r e^{i\phi}$, then the energy $E(u,v) = r$, and the measure $dudv = 2 \pi r dr$, so that 
\begin{eqnarray}
    D(E,\bk_0) &=& \frac{1}{{\rm det}[\hat e^a\cdot \vec\nabla Q_b]}
    \int \frac{2 \pi r dr}{(2\pi)^2}\delta (|E|- r)\cr
    &=& \frac{|E|}{2\pi {\rm det}[\hat e^a\cdot \vec\nabla Q_b]}
    \end{eqnarray}
so that 
\begin{equation}
D(E) = \frac{|E|}{2\pi}\int \frac{d^2 k_0}{(2\pi)^2}\frac{1}{{\rm det}[\hat e^a\cdot \vec\nabla Q_b]}
\end{equation}

\noindent demonstrating that the density of states $D(E)$ depdends linearly on energy. As a comparison with the analytical result,  we have computed the density of the states $D(E) = \frac{1}{\pi N}\sum_{k}\text{Im}[Tr(G_0(E-i0^+,\bk))]$ numerically as shown in Fig. \ref{DOS}(a) . A fit to the low energy density of states  (see Fig. \ref{DOS}(b)) confirms  $D(E)\propto |E|$. 
}

\begin{figure}[h]
   \includegraphics[width =7cm]{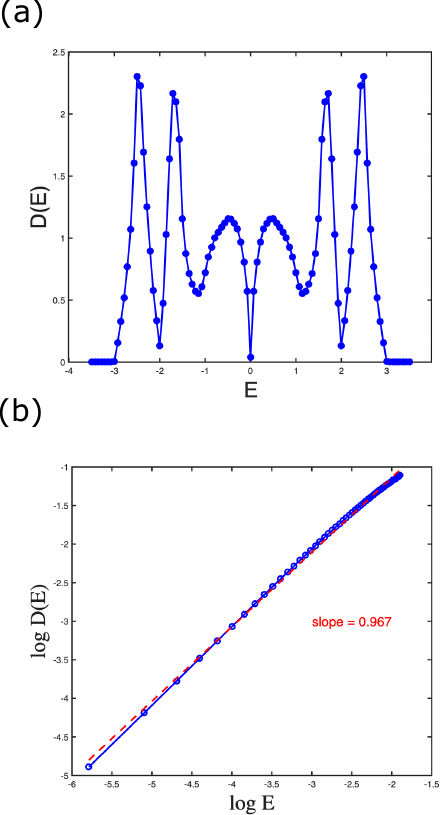}
   
   \caption{(a) The density of state as a function of energy. At zero energy, the density of states is zero, suggesting that the dimension of the Fermi surface is lower than three.  (b) Semilog plot for the DOS versus energy, with a slope that is consistent with $D(E)\propto |E|$ }\label{DOS}
\end{figure}

\subsection{The  ground state}

As in the 2D case, the ground-state excitations of the 4D Kitaev model maps onto a problem of free fermions, with diagonalized Hamiltonian
\begin{equation}
    H = \sum_{\bk \in \frac{1}{2}{\rm BZ},n\in [1,6]} E_{\bk n} d\dg_{\bk n} d_{\bk n}
\end{equation}
where $d^{\dagger}_{\bk,m}$ is the quasiparticle creation operator, $E_{\bk n}$ are the corresponding eigenstates and the sum runs over one-half of the Brillouin zone. We assume that the eigenstates are ordered, such that $E_{\bk n}<0$ for $n\in [1,3]$. The 
ground-state is then given by the filled Fermi sea
\begin{equation}
|GS \rangle = \prod_{\bk\in \frac{1}{2}{\rm BZ},n\in[1,3]}d^{\dagger}_{\bk n}|0\rangle .
\end{equation}
where the product over band-index is restricted to the first three negative energy quasiparticle states. 
The corresponding ground state energy is the summation of all negative energies 
\begin{equation}
E_{G} = \frac{1}{2}\sum_{n=1- 3}\int \frac{d^4k}{(2\pi)^4} E_{\bk n}.\end{equation} 
Evaluating the ground state energy numerically, we obtain $E_G = 
-2.29J$. 

\subsection{The anisotropic coupling constants}

We also identify the phase transition between the gapless phase and the gapped phase by tuning the anisotropy between the three coupling constants $J_x,J_y,J_z$ (see Fig. \ref{J_phase}). In contrast to the two-dimensional honeycomb lattice, where the equation for the gapless condition can be handled analytically, the parameter region that corresponds to gapped phases in our 4D model can only be numerically determined. To map out the phase diagram, we compute the minimum value of the gap between the eigenvalues $\Delta_g={\rm min}( E_4(\bk)-E_{3}(\bk))$ delineating the regions where $\Delta_g>0$ is finite. 

From the phase diagram, it is obvious that if the three coupling constants largely deviate from each other, the system is gapped, much similar to the case in the Kitaev honeycomb model\cite{KITAEV20062}. This result also suggests the possibility of realizing the topological order and non-abelian anyon in our model.
 \begin{figure}[H]
\centering
   \includegraphics[width =8cm]{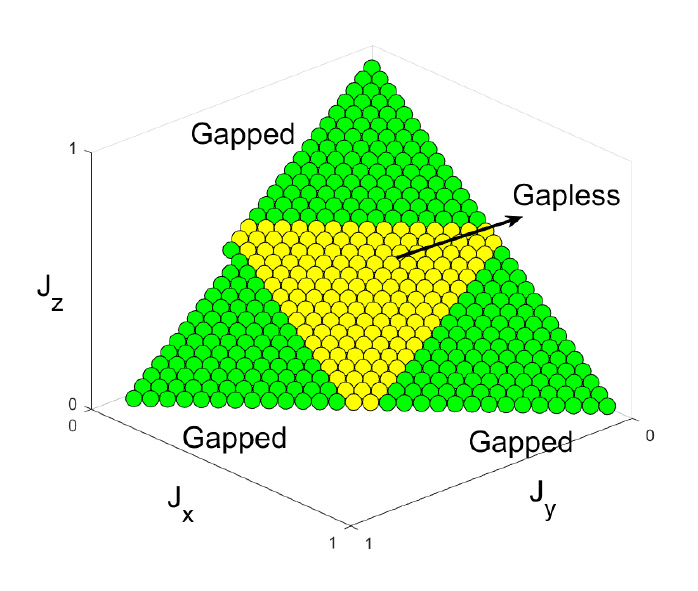}
   \caption{Gapped-Gapless phase diagram as a function of $J_x,J_y,J_z$ . The diagram looks very similar to the one for the 2D honeycomb model.}
   \label{J_phase}
\end{figure}

%

\section{Vison gap}\label{sec4}

 Although the energy spectrum and energy of the flux-free state are obtained in the previous section, one should notice that all the theoretical calculation done so far is based on the flux-free assumption. Central to our work is the assumption that the zero-flux state is energetically favorable. Although the Lieb theorem states that the ground state is flux-free for a two-dimensional system, the theorem only applies to two dimensions. To confirm that the vortex-free configuration is the lowest energy state in our model, we need to confirm that the energy of single bond-flip is positive, allowing stability against
 gauge fluctuations.

 \begin{figure}[H]
\centering
   \includegraphics[width =8cm]{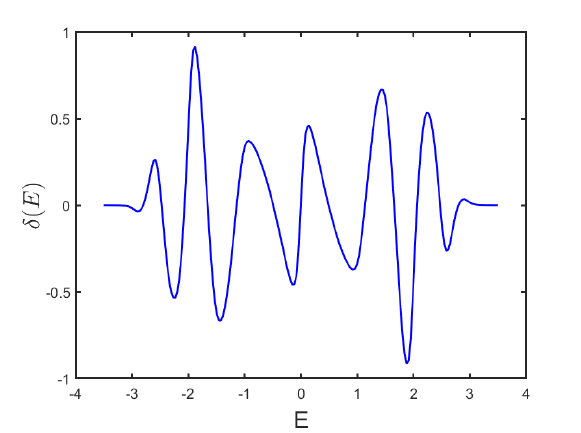}
   \caption{Energy dependence of the scattering phase shift resulting from bond-flipping in the hyper{-}hexagonal Kitaev model. The energy is expressed in the unit of $J$.}
   \label{phase_shift}
\end{figure}

To carry out this calculation, we follow the methodology of \cite{PhysRevB.108.045151}, evaluating the energy cost of a bond-flip in terms of the Majorana scattering phase shift. 
 For simplicity, we consider the isotropic case($J_x=J_y=J_z=J$). The formalism is summarized in the Appendix \ref{AppE} . Here we only mention the crucial results. The presence of the bond flipping potential causes the phase shift and the energy change due to the vison creation is given by the energy integration of the phase shift  $\Delta F = \int_0^{\infty} dE \frac{\delta(E)}{2\pi}$, where $\delta(E)$ is the scattering phase shift. The FIG. \ref{phase_shift} shows our numerical calculation of the $\delta(E)$ in the case where the bond variable between $3^{rd}$ and $4^{th}$ atoms $u_{\langle34\rangle}$ is flipped. After numerical integration, we find that the change of energy due to the bond flip is positive, given by  $0.0234 (J)$ . This numerical calculation confirms that the flux-free state is locally stable against gauge fluctuations.

\section{Discussion and Conclusion}\label{sec5}

Here we discuss some of the forward implications of our work and some of the open questions it poses.

{ \subsection{Fractionalization and the Yao-Lee extension}

The factorization of the spin operator 
\begin{equation}
    \vec \sigma_j = 2i a_j \vec b_j
\end{equation}
into a bilinear of free fermions, together with the 
positive vison energy computed in section \ref{sec4} is a strong indicator of fractionalization in our 4D Kitaev spin liquid.  As in lower dimensional spin liquids, a spin-flip creates a mobile Majorana while also creating two visons, thus a strict statement about the fractionalization requires some knowledge of the statistical mechanics of the static $Z_2$ gauge fields. 
For example, confirmation that the spin-susceptibilty involves a continuum of excitations above the vison gap requires a study of the vison correlation functions, which requires numerical methods\cite{PhysRevB.92.115127}.  In two dimensions, the separation of Majorana and gauge degrees of freedom can also be confirmed by showing that the specific heat curves separate into a low-temperature  Majorana component and a high temperature gauge component, each carrying entropy $\frac{1}{2}\ln 2$ per site\cite{PhysRevB.92.115122}.

However, in three and higher dimensions, fractionalization is guaranteed by the properties of the $Z_2$ gauge fields. When the fermions are integrated out of the partition function, the remaining effective action describes a pure $Z_2$ lattice gauge theory, which is subject to a finite temperature Ising transition between a high temperature area-law phase and a low temperature perimeter law phase in which the visons are confined \cite{wegnerDualityGeneralizedIsing1971,PhysRevD.19.3682}.  The presence of this transition in ($D>2$)  Kitaev models has been explicitly confirmed without performing a heavy numerical calculation. Below the Ising transition temperature, the Majorana fermion $a_j$ becomes a free fermion, guaranteeing a fractionalized phase. 

An even clearer rendition of this fractionalization can be obtained by constructing the corresponding Yao Lee model on the hyper-hexagonal lattice. Here we can identify the formation of a Yao-Lee spin liquid(YLSL) with gapless, fractionalized spin excitations\cite{PhysRevLett.107.087205,PhysRevB.106.125144}. The  Yao–Lee generalization of a  Kitaev model incorporates an additional spin-1/2 orbital degree of freedom $\lambda_j^{\alpha}$ ($\alpha = 1-3$), in addition to a spin degree of freedom $\vec\sigma_j$ at each site, where the $\vec \lambda_j$ and $\vec \sigma_j$ are independent Pauli matrices.  The Yao-Lee Hamiltonian 
\begin{equation}\label{YL_model}
    \begin{split}
        H = J\sum_{\langle i,j\rangle}\lambda_i^{\alpha_i}
        \lambda_j^{\alpha_j}(\vec \sigma_i\cdot\vec \sigma_j).
    \end{split}
\end{equation}
involves a frustrated  Ising coupling between the orbital fields, decorated by a Heisenberg interaction between the spin fields. In this model, the factorization of Pauli operators involves defining a triplet of Majorana fermions $\vec b_j = (b_j^1,b_j^2,b_j^3)$ and $\vec \chi_j = (\chi_j^1,\chi_j^2,\chi_j^3)$\cite{PhysRevB.106.125144}. Within this representation, the Pauli operators can be written as 
\begin{equation}
    \begin{split}
    \vec \sigma_j=-i\vec\chi_j\times\vec\chi_j,\\
        \vec \lambda_j=-i\vec b_j\times\vec b_j.
    \end{split}
\end{equation}
After the substitution, the Hamiltonian can be expressed as 
\begin{equation}
         H = J\sum_{\langle i,j\rangle}\hat u_{ij}(i\vec\chi_i\cdot\vec\chi_j),
\end{equation}
where $\hat u_{ij}=-2ib^{\alpha_i}_ib^{\alpha_j}_j$. One can observe that the gauge degree of freedom is carried solely by the orbital Majorana fermions $\vec b_j$ and is decoupled from spin components. Therefore, a spin-flip no longer involves the gauge field and the creation of visons.   Thus in the Yao-Lee extension of our model, we can ensure that the spin fractionalizes into gapless particles, and that the dynamical spin susceptibility acquires a gapless Lindhard continuum associated with the 2D Majorana Fermi surface.}

\subsection{Jordan-Wigner transformation}

We can also derive the same results using the Jordan-Wigner  transformation\cite{Jordan1928}. This approach can be applied to the Kitaev model and its variants, as discussed in Ref.\cite{Jin2023,Feng2007,Chen2008} . The string operator, which enforces the commutivity of spins at different sites, is an essential element of the  Jordan-Wigner transformation and can be consistently defined with  open boundary conditions.  Usually, a Jordan-Wigner transformation  in higher dimensional models gives rise to long non-local strings. Fortunately, a Kitaev-type model eliminates this issue, since the bonds separating the strings are  $z$ bonds, which create inter-chain gauge fields that do  not interfere with the locality of the  model. The construction of the string operator relies on definition of a ordered path. 
{
The path is defined as follows: the path first advances monotonically along the spiral direction, sweeping through all sites from one end of the system to the other in each spatial dimension. Then the Jordan-Wigner transform for spin operators can be expressed as
\begin{equation}
    \begin{split}
       S^{+}_i &= c^\dagger_iP_i \\
       S^{-}_i &= c_iP_i \\
       S^z_i &= c^\dagger_ic_i - 1,
    \end{split}
\end{equation}
where $c$ denotes spinless fermion operators. $P_i = e^{i\Phi}$,with $\Phi = \pi \sum_{j<i}n_{j}$ ,is the string operators running according to the ordered path as defined above.} To proceed, we perform the Jordan-Wigner transform and express the complex fermion operators as Majorana operators. The final Hamiltonian turns out to be
\begin{equation}
    \begin{split}
        H &= \sum_{<ij>\in x-bond} J_x  c_{A,i}c_{A,j}+
        \sum_{<ij>\in y-bond} J_y  c_{A,i}c_{A,j}+\\
        &\sum_{<ij>\in z-bond} J_z D_{ij} c_{A,i}c_{A,j},
    \end{split}
\end{equation}
where the $c_{\alpha,i},\alpha=A,B$ are Majorana operators. The operator $D_{ij} = \chi_{B,i}\chi_{B,j}$ has an eigenvalue $\pm 1$. Since each $D_{ij}$ lives on a $z$ bond and does not overlap with each other, this operator commutes with the Hamiltonian and becomes a constant of motion. This then reverts to Hamiltonian \eqref{theham}, but in the gauge where the $u_{ij}$ are restricted to lie along the $z$ bonds between the chains. In this way, we see that the Jordan-Wigner transformation works in four dimensions.

{
\subsection{Finite Zeeman Field }

\begin{figure}[t]
\centering
   \includegraphics[width =8cm]{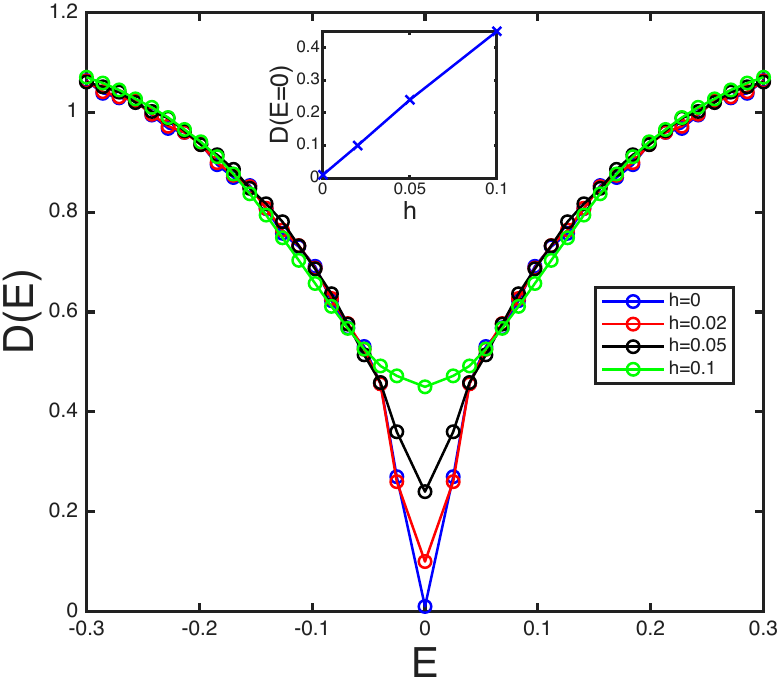}
   \caption{Density of state as a function of energy. The three curve correspond to three different field strength $h = \frac{h_xh_yh_z}{J^2}=0,0.02,0.05,0.1$. The inset plot shows the density of state at Fermi energy $E=0$ as a function of field strength.}
   \label{DOS_h}
\end{figure}

 In addition, based on our 4D model, one can ask how the system behaves in the presence of a magnetic field. In the Kitaev honeycomb model\cite{KITAEV20062}, a magnetic Zeeman coupling induces a next-nearest neighboring hopping term which leads to a topological gap opening and formation of a Chern insulator, in analogy to the Haldane model\cite{PhysRevLett.61.2015}. In principle,  
 four dimension systems allow for the existence of non-Abelian Chern insulators characterized by the second Chern number $C_2$, provided the field continues to induce a gap. 
 In the presence of a Zeeman coupling $- {\bf h} \cdot \mathbf{\sigma_j}$ at each site, 
 the Majorana Hamiltonian develops next-nearest neigbor hopping terms, giving rise to a Hamiltonian of the form $H = \sum_{\bk}\tilde{c}\dg_{\bf k} \bigl[{\cal H}({\bf k} ) + \Lambda ({\bf k})\bigr]\tilde c_{\bf k}$ where ${\cal H}({\bf k})$ is given in \eqref{HQ} and
 \begin{equation}
 {\Lambda }(\bk) = \pmat{F(\bk) & 0\cr 0 & -G(\bk)}.
  \end{equation}
  has a block-diagonal structure with 
  \begin{equation}
  F(\bk) = -\alpha \frac{h_xh_yh_z}{J^2}\pmat{0 & - i f_{13}& i f_{15}\cr
  i f^*_{13} & 0 & - i f_{35} \cr
   -if^*_{15}& i f^*_{35} & 0}
  \end{equation}
  and
  \begin{equation}
  G(\bk) = -\alpha \frac{h_xh_yh_z}{J^2}\pmat{0 & - i f_{24}& i f_{26}\cr
  i f^*_{24} & 0 & - i f_{46} \cr
   -if^*_{26}& i f^*_{46} & 0}
  \end{equation}
  where $\alpha \sim O(1)$ is a number of order unity, $h_{x},h_y, h_z$ are the components of the applied field and 
  \begin{eqnarray}
      f_{13}&=&1+e^{-i {k_1}}+e^{i ({k_3}-{k_4})},
\cr
f_{15}&=&e^{-i {k_1}}+e^{i {k_2}}+e^{-i {k_4}},
\cr
f_{35}&=&1+e^{i {k_2}}+e^{-i {k_3}},
  \end{eqnarray}
  and
  \begin{eqnarray}
      f_{24}&=&1+e^{-i {k_1}}+e^{i {k_2}},
\cr
f_{26}&=&e^{i {k_2}}+e^{-i {k_3}}+e^{-i {k_4}},
\cr
f_{46}&=&1+e^{i ({k_1}-{k_4})}+e^{-i {k_3}}
  \end{eqnarray}
are the form-factors for next nearest neighbor hopping. 
  
In a 2D Kitaev model, ${\cal H}(\bk) + \Lambda(\bk)  = \vec d(\bk)\cdot \vec \sigma$ can be decomposed in a set of anti-commuting Pauli matrices, giving rise to a dispersion $E_{\bk} = \pm |\vec d(\bk)|$.
When there are only two components to $\vec d(\bk)$, the condition that $E(\bk)=0$ defines two constraints, defining the two gapless Dirac points, but in a magnetic field, the condition that three components vanish cannot be satisfied anywhere in momentum space, leading to a gapped spectrum. By contrast, in the 4D Kitaev model, $\{{\cal H}(\bk), \Lambda(\bk) \}\neq 0$ and the complex-valued condition $\det Q(\bk)=Q_1(\bk) + i Q_2(\bk)=0 $ which defines a 2D Fermi surface with a vanishing density of states in zero field becomes a single, real-valued condition ${\rm det}[{\cal H}(\bk)+ \Lambda(\bk)]=0$ in a field. This condition now defines a { 3D } Fermi surface. In this situation, the spin-liquid in a Zeeman field develops a {\sl  finite } density of states. Numerical  calculations shown in Fig. \ref{DOS_h} confirm this expectation, showing the development of a finite density of states proportional to $h_xh_yh_z$. This rules out the formation of a topological gapped state in our model.

 }

 \subsection{Conclusion}
To conclude, we have proposed an exactly solvable four-dimensional Kitaev model. The key ingredient is a spiral structure which allows a well-defined Jordan-Wigner string. Our model calculation demonstrates that the 4D Kitaev spin model possesses a two-dimensional Majorana Fermi surface. We have  used  scattering theory to compute the energy cost to create a vison, demonstrating that the state with zero-flux is  locally stable. Furthermore, we also examine the effect of the Zeeman field. Contrary to expectations, our model develops a three-dimensional Fermi surface in a field, with a finite density of states proportional to the cube of the applied fields. 
{
Our 4D model opens a new venue for exploring high dimensional topological phases and inspires several open questions. For instance, one can ask whether our model can be extended to higher dimensions.  Clearly the spiral motif can be extended to higher dimensions, but without further work, it is not yet clear that the resulting structures
will close to produce a consistent crystal structure. 

Another challenge for future work is to further explore whether a fully gapped non-Abelian phase can be found in a four-dimension generalization of a Kitaev model.  For this purpose, it might be useful to employ a higher-dimensional Clifford algebra which allows lattices with larger co-ordination numbers, as discussed in Ref. \cite{PhysRevB.102.201111}.}

\vskip 0.1in

\section*{Data Availability}

The datasets generated and analyzed during the current study, together with the codes for calculating the density of states and the phase shift are available
in Ref.~\cite{mydata2026}.

\begin{acknowledgments}
	
		We would like to thank Aaditya Panigrahi for fruitful discussions. This work was supported by Office of Basic Energy Sciences, Material
		Sciences and Engineering Division, U.S. Department of Energy (DOE)
		under Contract DE-FG02-99ER45790.

\end{acknowledgments}

\onecolumngrid

\appendix
\counterwithin*{equation}{section} 
\renewcommand\theequation{\thesection\arabic{equation}} 
\section{The hyper-hexagonal lattice}\label{AppA}

\begin{figure}[H]
    \centering
    \includegraphics[width=0.8\linewidth]{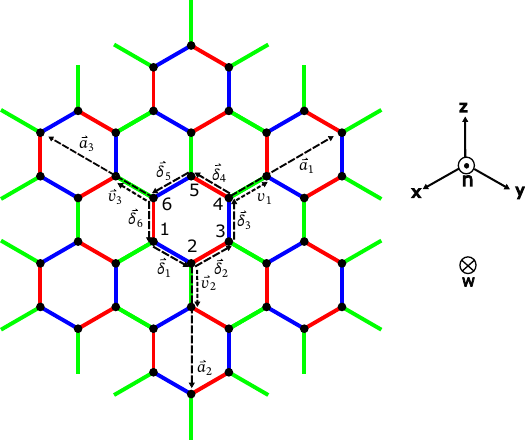}
    \caption{The projected 2D lattice structure of the hyper-hexagonal lattice. The projection direction is $\hat n = (1,1,1,0)$. The blue-red bonds denote $xx,yy$ alternative bonds, while the green bonds represent $zz$ bonds.}
    \label{2D_lattice}
\end{figure}

Here, we provide a detailed description of the hyper-hexagonal lattice. All vectors used in the following calculation are depicted in  FIG. \ref{2D_lattice}. Recall the vectors connecting different sites inside a unit cell

\begin{equation}
    \begin{split}
        \boldsymbol{\delta_1} &= (0,1,0,1),\\
        \boldsymbol{\delta_2} &= (-1,0,0,0),\\
        \boldsymbol{\delta_3} &= (0,0,1,1),\\
        \boldsymbol{\delta_4} &=(0,-1,0,0),\\
        \boldsymbol{\delta_5} &= (1,0,0,1),\\
        \boldsymbol{\delta_6} &= (0,0,-1,0),\\
    \end{split}
\end{equation}
where only $\bm{\delta_1},\bm{\delta_3}$and $\bm{\delta_5}$ have a non-vanishing fourth component. We also define the linking vectors $\bm{v_1},\bm{v_2}$, and $\bm{v_3}$ connecting the neighboring unit cells. 
\begin{equation}
    \begin{split}
        \boldsymbol{v_1} &= \frac{1}{2}(-1,1,1,0),\\
        \boldsymbol{v_2} &= \frac{1}{2}(1,1,-1,0),\\
          \boldsymbol{v_3} &= \frac{1}{2}(1,-1,1,0).\\
    \end{split}
\end{equation}
 Our choice of vector follows the rule that the lengths of vectors $\bm{v_i}$ are the same as those of red and blue bonds. This setup allows us to define three Bravais lattice vectors characterizing the periodicity of the lattice geometry

\begin{equation}
    \begin{split}
        \boldsymbol{a}_1 &= \boldsymbol{\delta}_1+\boldsymbol{\delta}_2 + \boldsymbol{\delta}_3 + \boldsymbol{v}_1
        =(-\frac{3}{2},\frac{3}{2},\frac{3}{2},2)\\
        \boldsymbol{a}_2 &= \boldsymbol{\delta}_2+
\boldsymbol{\delta}_3 + \boldsymbol{\delta}_4 + \boldsymbol{v}_2
=(\frac{3}{2},\frac{3}{2},-\frac{3}{2},-1)\\
        \boldsymbol{a}_3 &= \boldsymbol{\delta}_3+
\boldsymbol{\delta}_4+
\boldsymbol{\delta}_5 + \boldsymbol{v}_3
=(\frac{3}{2},-\frac{3}{2},\frac{3}{2},2)\\
        \boldsymbol{a}_4 &= \boldsymbol{\delta}_1
        +\boldsymbol{\delta}_2
        +\boldsymbol{\delta}_3
        +\boldsymbol{\delta}_4
        +\boldsymbol{\delta}_5
        +\boldsymbol{\delta}_6
        =(0,0,0,3)
    \end{split}
\end{equation}

Next, we will show that the lattice geometry in a two-dimensional projected space is identical to the honeycomb lattice structure. To begin with, we aim to obtain vectors that describe the lattice geometry in a two-dimensional projected space. To do this, we first project out the fourth component(the projection direction is $\bm{w}=(0,0,0,1)$). The second step is to project out a particular direction in the 3D space. Thus, we introduce the normal vector $\boldsymbol{n} = \frac{1}{\sqrt{3}}(1,1,1,0)$. We then define an operator for the two-step projection
\begin{equation}
    \begin{split}
        P&= I_{4\times4}-\boldsymbol{n}\boldsymbol{n}^T-\bm{w}^T\bm{w}\\
        &= 
        \frac{1}{3}\left(
        \begin{array}{cccc}
            2 & -1 & -1 & 0\\
             -1&2  &-1& 0 \\
             -1&-1 &2& 0\\
              0& 0& 0& 0
        \end{array}
        \right).
    \end{split}
\end{equation}

    The projected vectors are then
\begin{equation}
    \begin{split}
        \boldsymbol{\delta}_{1,2D} = P\boldsymbol{\delta}_1 &= (-\frac{1}{3},\frac{2}{3},-\frac{1}{3},0)^T\\
        \boldsymbol{\delta}_{2,2D} = P\boldsymbol{\delta}_2 &= (-\frac{2}{3},\frac{1}{3},\frac{1}{3},0)^T\\
        \boldsymbol{\delta}_{3,2D} = P\boldsymbol{\delta}_3 &= (\frac{-1}{3},\frac{-1}{3},\frac{2}{3},0)^T\\
         \boldsymbol{\delta}_{4,2D} = P\boldsymbol{\delta}_4 &= (\frac{1}{3},-\frac{2}{3},\frac{1}{3},0)^T\\
        \boldsymbol{\delta}_{5,2D} = P\boldsymbol{\delta}_5 &= (\frac{2}{3},-\frac{1}{3},-\frac{1}{3},0)^T\\
        \boldsymbol{\delta}_{6,2D} = P\boldsymbol{\delta}_6 &= (\frac{1}{3},\frac{1}{3},-\frac{2}{3},0)^T\\
        \boldsymbol{v}_{1,2D} = P\boldsymbol{v}_1 &= (-\frac{2}{3},\frac{1}{3},\frac{1}{3},0)^T\\
         \boldsymbol{v}_{2,2D} = P\boldsymbol{v}_2 &= (\frac{1}{3},\frac{1}{3},-\frac{2}{3},0)^T\\
          \boldsymbol{v}_{3,2D} = P\boldsymbol{v}_3 &= (\frac{1}{3},-\frac{2}{3},\frac{1}{3},0)^T\\
    \end{split}
\end{equation}

 Since we have projected out two degrees of freedom, we can define two basis vector $\bm{\hat x}=\frac{1}{\sqrt{2}}(-1,1,0,0)$ and $\bm{\hat y}=\frac{1}{\sqrt{6}}(-1,-1,2,0)$
to re-express the above vectors. A simple algebra leads to the following results

\begin{equation}
\begin{split}
        \boldsymbol{\delta}_{1,2D}&
    = \frac{1}{\sqrt{2}}\bm{\hat x}-\frac{1}{\sqrt{6}}\bm{\hat y},
    \boldsymbol{\delta}_{2,2D}
    = \frac{1}{\sqrt{2}}\bm{\hat x}+\frac{1}{\sqrt{6}}\bm{\hat y}\\
    \boldsymbol{\delta}_{3,2D}& 
    = \sqrt{\frac{2}{3}}\bm{\hat y},
    \boldsymbol{\delta}_{4,2D} 
    =-\frac{1}{\sqrt{2}}\bm{\hat x}+\frac{1}{\sqrt{6}}\bm{\hat y},\\
    \boldsymbol{\delta}_{5,2D}& 
    =-\frac{1}{\sqrt{2}}\bm{\hat x}-\frac{1}{\sqrt{6}}\bm{\hat y},
    \boldsymbol{\delta}_{6,2D} 
    = -\sqrt{\frac{2}{3}}\bm{\hat y}\\
    \boldsymbol{v}_{1,2D}&=
    \frac{1}{\sqrt{2}}\bm{\hat x}+\frac{1}{\sqrt{6}}\bm{\hat y}\\
    \boldsymbol{v}_{2,2D}&= -\sqrt{\frac{2}{3}}\bm{\hat y}\\
    \boldsymbol{v}_{3,2D}&= 
     -\frac{1}{\sqrt{2}}\bm{\hat x}+\frac{1}{\sqrt{6}}\bm{\hat y}
\end{split}
\end{equation}
It is easy to prove that all these vectors can form a hexagonal structure in a 2D space as shown in Fig. \ref{2D_lattice}.

\section{Scattering theory}\label{AppE}
As mentioned in the main text, the flux-free state is not necessarily the ground state in our model since the validity of  Lieb theorem\cite{Lieb1994} is not guaranteed in a higher-dimensional system. To examine whether the flux-free state is the ground state, we consider a scenario where the {visons} are created via flipping a gauge-embedded bond. We then adapt the scattering theoretic description, treating the bond flip as a scattering potential. Subsequently, we utilize the Green function approach \cite{PhysRevB.108.045151,Panigrahi2024} to evaluate the change of free energy due to flipping a bond on the lattice. Let us consider the case of flipping a single bond $u_{ij}$ for a $xx$ or a $yy$ bond. We then rewrite the Hamiltonian as a flux-free part and bond-flip part, $H = H_{0} + V$, where
\begin{equation}
    \begin{split}
        H_0  &= \sum_{i,j}2iJ_{\alpha_{ij}}u_{\langle ij\rangle}c_{i}c_{j},\text{with all $u_{\langle ij\rangle}=1$},\\
        V &= -4J_{\alpha_{ab}}ic_{a}c_{b}.
    \end{split}
\end{equation}
Here, the index $a,b$ denote two adjacent atoms within the unit cell. To calculate the vison gap, we first write down the free energy as
\begin{equation}
    \begin{split}
        F = \frac{-1}{2\beta}\sum_{\omega,k}\mathrm{tr}(\ln(-G^{-1})),
    \end{split}
\end{equation}
where $\beta = 1/T$. The factor $\frac{1}{2}$ on the right-hand side prevents the double counting of modes in the Brillouin zone. The trace denotes the summation over the sublattice degree of freedom. $G$ is the Green function, which can be expressed as $G = (G_{0}^{-1}-V)^{-1}$ . $G_{0}$ is  a flux-free Green function given by
\begin{equation}
    \begin{split}
        G_{0}(i\omega,\bk) = \frac{1}{i\omega - H_0(\bk)}= \sum_{n}\frac{|\psi_{n}\rangle \langle \psi_{n}|}
        {i\omega-\epsilon_{n}(\bk)}.
    \end{split}
\end{equation}
$\epsilon_n(k)$ represents the eigen-energy of the Hamiltonian $H_0(k)$ and $|\Psi_n\rangle$ is the corresponding eigenstate. Separating zero-flux free energy and change of free energy due to scattering potential, it yields

\begin{equation}
    \begin{split}
        \Delta F = -\frac{-1}{2\beta}\sum_{i\omega_n}tr(1-\hat V \sum_{k}G_{0}(i\omega,\bk)).
    \end{split}
\end{equation}

For our convenience, one can define the local Green function
\begin{equation}
    \begin{split}
        g(i\omega_{n}) = \sum_{k}G_{0}(i\omega_n,\bk).
    \end{split}
\end{equation}
Thus the change of free energy becomes 
\begin{equation}
    \begin{split}
        \Delta F = \frac{-1}{2\beta} \sum_{i\omega}tr(1-\hat V g(i\omega_n)).
    \end{split}
\end{equation}

Additionally, the bond-flipping potential can be expressed in matrix form
 \begin{equation}
     \begin{split}
         \hat V = V O_{m}\oplus\sigma_2\oplus O_{4-m},
     \end{split}
 \end{equation}
where V is the coupling constant and $\sigma_2$ is the Pauli-y matrix. $O_m$ is a $m$ by $m$  null matrix, and the index $m$ denotes bond flipping between $m+1^{th}$ and $m+2^{th}$ atoms. As noted in the maintext, we consider the case where bond flipping occurs at the bond connecting $3^{rd}$ and $4^{th}$ atoms,or equivalently, $m=2$. Therefore, the final expression of the free energy is
\begin{equation}
    \begin{split}
        \Delta F = \frac{-1}{2\beta} \sum_{i\omega_n}tr(\ln(1- V O_{2}\oplus\sigma_2\oplus O_{2} g(i\omega_n))).
    \end{split}
\end{equation}
The trace of the logarithm function can be further simplified as the logarithm of the determinant of the matrix. It yields
\begin{equation}
    \begin{split}
        \Delta F = \frac{-1}{2\beta}\sum_{i\omega_n,\alpha}\ln \lambda_{\alpha} (i\omega_n),
    \end{split}
\end{equation}
where $\lambda_{\alpha}$ is an eigenvalue of $1-\hat V O_{2}\oplus\sigma_2\oplus O_{2} g(i\omega_n)$. We can further evaluate the Matsubara summation by converting the summation into a real energy integral, see Fig. \ref{contour}.
 \begin{figure}[H]
\centering
   \includegraphics[width =8cm]{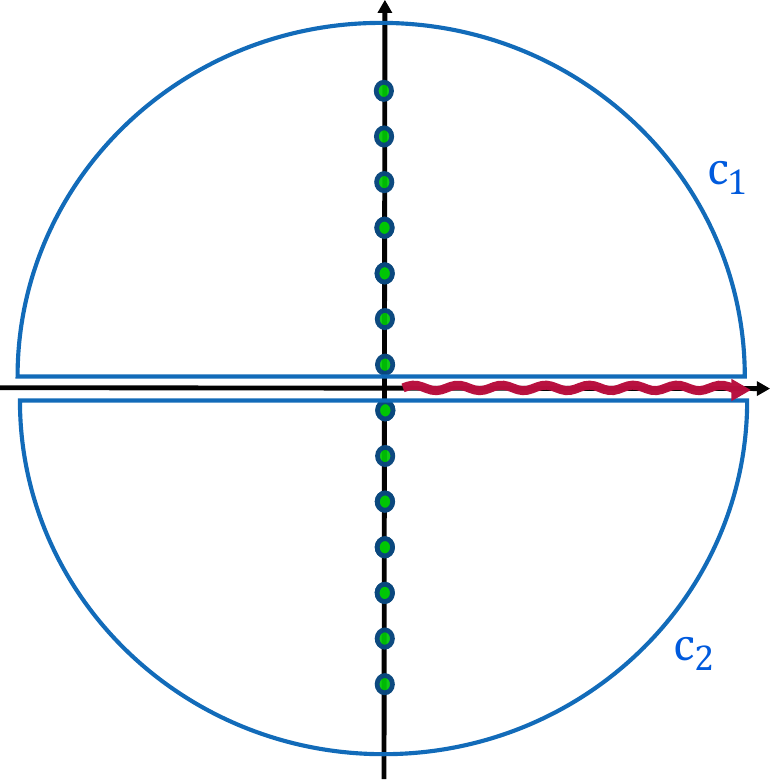}
   \caption{The contour $C_1,C_2$ for a known branch cut. }
   \label{contour}
\end{figure}
\begin{equation}
    \begin{split}
        \Delta F &= -\int^{\infty
}_{-\infty} \frac{d\omega}{2\pi} f(\omega) Im 
  \ln\lambda_{\alpha}(\omega)\\
  & = -\int^{\infty
}_{-\infty} \frac{d\omega}{2\pi} f(\omega) \delta (\omega).
    \end{split}
\end{equation}
Here, $\delta(\omega)$ denotes the scattering phase shift. Since the phase shift is odd in frequency ($\delta(-\omega) = -\delta(\omega)$). At zero temperature, the integral can be re-expressed 
\begin{equation}
    \Delta F = \int^{\infty}_{0} \frac{d\omega}{2\pi}\delta(\omega).
\end{equation}
The remaining part of the calculation is performed numerically. Performing the numerical integration, one can obtain that the single bond flip energy is around 0.0234(J). The energy cost to create visons is positive, indicating that the flux-free state has the lowest energy among all possible gauge configurations. Another issue is the energy cost to flip a $zz$ bond. The scattering potential would be non-local since the two points forming $zz$ bonds belong to different unit cells, according to our definition. A straightforward symmetry argument tells us the energy cost to flip a $zz$ bond is different from that of flipping $xx,yy$ bonds.   

\twocolumngrid

\bibliography{ref}

@article{Anderson1987,
author = {P. W. Anderson },
title = {{The Resonating Valence Bond State in La$_2$CuO$_4$ and Superconductivity}},
journal = {Science},
volume = {235},
number = {4793},
pages = {1196-1198},
year = {1987},
doi = {10.1126/science.235.4793.1196}}

@misc{mydata2026,
  author       = {Chen, Po-Jui and Coleman, Piers},
  title        = {Dataset for ``A higher dimensional generalization of the Kitaev spin liquid (https://doi.org/10.5281/zenodo.19199451)''},
  year         = {2026},
  publisher    = {Zenodo},
  doi          = {
  10.5281/zenodo.19199451},
  url          = {https://doi.org/10.5281/zenodo.19199451}
}

@article{PhysRevB.108.045151,
  title = {{Analytic calculation of the vison gap in the Kitaev spin liquid}},
  author = {Panigrahi, Aaditya and Coleman, Piers and Tsvelik, Alexei},
  journal = {Phys. Rev. B},
  volume = {108},
  issue = {4},
  pages = {045151},
  numpages = {7},
  year = {2023},
  month = {Jul},
  publisher = {American Physical Society},
  doi = {10.1103/PhysRevB.108.045151},
  url = {https://link.aps.org/doi/10.1103/PhysRevB.108.045151}
}

@article{Savary2016,
doi = {10.1088/0034-4885/80/1/016502},
url = {https://dx.doi.org/10.1088/0034-4885/80/1/016502},
year = {2016},
month = {nov},
publisher = {IOP Publishing},
volume = {80},
number = {1},
pages = {016502},
author = {Lucile Savary and Leon Balents},
title = {{Quantum spin liquids: a review}},
journal = {Reports on Progress in Physics}
}

@article{balents_spin_2010,
	title = {{Spin liquids in frustrated magnets}},
	volume = {464},
	issn = {0028-0836, 1476-4687},
	url = {https://www.nature.com/articles/nature08917},
	doi = {10.1038/nature08917},
	number = {7286},
	urldate = {2024-10-20},
	journal = {Nature},
	author = {Balents, Leon},
	month = mar,
	year = {2010},
	pages = {199--208} }

@article{KITAEV20062,
title = {{Anyons in an exactly solved model and beyond}},
journal = {Annals of Physics},
volume = {321},
number = {1},
pages = {2-111},
year = {2006},
note = {January Special Issue},
issn = {0003-4916},
doi = {https://doi.org/10.1016/j.aop.2005.10.005},
url = {https://www.sciencedirect.com/science/article/pii/S0003491605002381},
author = {Alexei Kitaev}
}

@article{Eschmann2020,
  title = {{Thermodynamic classification of three-dimensional Kitaev spin liquids}},
  author = {Eschmann, Tim and Mishchenko, Petr A. and O'Brien, Kevin and Bojesen, Troels A. and Kato, Yasuyuki and Hermanns, Maria and Motome, Yukitoshi and Trebst, Simon},
  journal = {Phys. Rev. B},
  volume = {102},
  issue = {7},
  pages = {075125},
  numpages = {24},
  year = {2020},
  month = {Aug},
  publisher = {American Physical Society},
  doi = {10.1103/PhysRevB.102.075125},
  url = {https://link.aps.org/doi/10.1103/PhysRevB.102.075125}
}

@article{Anderson1973,
title = {{Resonating valence bonds: A new kind of insulator?}},
journal = {Materials Research Bulletin},
volume = {8},
number = {2},
pages = {153-160},
year = {1973},
issn = {0025-5408},
doi = {https://doi.org/10.1016/0025-5408(73)90167-0},
url = {https://www.sciencedirect.com/science/article/pii/0025540873901670},
author = {P.W. Anderson}
}

@article{Feng2007,
  title = {{Topological Characterization of Quantum Phase Transitions in a Spin-$1/2$ Model}},
  author = {Feng, Xiao-Yong and Zhang, Guang-Ming and Xiang, Tao},
  journal = {Phys. Rev. Lett.},
  volume = {98},
  issue = {8},
  pages = {087204},
  numpages = {4},
  year = {2007},
  month = {Feb},
  publisher = {American Physical Society},
  doi = {10.1103/PhysRevLett.98.087204},
  url = {https://link.aps.org/doi/10.1103/PhysRevLett.98.087204}
}

@article{Jahromi2021,
  title = {{Thermodynamics of three-dimensional Kitaev quantum spin liquids via tensor networks}},
  author = {Jahromi, Saeed S. and Yarloo, Hadi and Or\'us, Rom\'an},
  journal = {Phys. Rev. Res.},
  volume = {3},
  issue = {3},
  pages = {033205},
  numpages = {15},
  year = {2021},
  month = {Sep},
  publisher = {American Physical Society},
  doi = {10.1103/PhysRevResearch.3.033205},
  url = {https://link.aps.org/doi/10.1103/PhysRevResearch.3.033205}
}

@article{Lieb1994,
  title = {{Flux Phase of the Half-Filled Band}},
  author = {Lieb, Elliott H.},
  journal = {Phys. Rev. Lett.},
  volume = {73},
  issue = {16},
  pages = {2158--2161},
  numpages = {0},
  year = {1994},
  month = {Oct},
  publisher = {American Physical Society},
  doi = {10.1103/PhysRevLett.73.2158},
  url = {https://link.aps.org/doi/10.1103/PhysRevLett.73.2158}
}

@article{Hermanns2014,
  title = {Quantum spin liquid with a Majorana Fermi surface on the three-dimensional hyperoctagon lattice},
  author = {Hermanns, M. and Trebst, S.},
  journal = {Phys. Rev. B},
  volume = {89},
  issue = {23},
  pages = {235102},
  numpages = {16},
  year = {2014},
  month = {Jun},
  publisher = {American Physical Society},
  doi = {10.1103/PhysRevB.89.235102},
  url = {https://link.aps.org/doi/10.1103/PhysRevB.89.235102}
}

@article{Yamada2017,
  title = {Crystalline Kitaev spin liquids},
  author = {Yamada, Masahiko G. and Dwivedi, Vatsal and Hermanns, Maria},
  journal = {Phys. Rev. B},
  volume = {96},
  issue = {15},
  pages = {155107},
  numpages = {15},
  year = {2017},
  month = {Oct},
  publisher = {American Physical Society},
  doi = {10.1103/PhysRevB.96.155107},
  url = {https://link.aps.org/doi/10.1103/PhysRevB.96.155107}
}

@article{Hermanns2015,
  title = {Weyl Spin Liquids},
  author = {Hermanns, M. and O'Brien, K. and Trebst, S.},
  journal = {Phys. Rev. Lett.},
  volume = {114},
  issue = {15},
  pages = {157202},
  numpages = {5},
  year = {2015},
  month = {Apr},
  publisher = {American Physical Society},
  doi = {10.1103/PhysRevLett.114.157202},
  url = {https://link.aps.org/doi/10.1103/PhysRevLett.114.157202}
}

@article{Tsvelik2022,
  title = {{Order fractionalization in a Kitaev-Kondo model}},
  author = {Tsvelik, Alexei M. and Coleman, Piers},
  journal = {Phys. Rev. B},
  volume = {106},
  issue = {12},
  pages = {125144},
  numpages = {16},
  year = {2022},
  month = {Sep},
  publisher = {American Physical Society},
  doi = {10.1103/PhysRevB.106.125144},
  url = {https://link.aps.org/doi/10.1103/PhysRevB.106.125144}
}

@article{Coleman2022,
  title = {{Solvable 3D Kondo Lattice Exhibiting Pair Density Wave, Odd-Frequency Pairing, and Order Fractionalization}},
  author = {Coleman, Piers and Panigrahi, Aaditya and Tsvelik, Alexei},
  journal = {Phys. Rev. Lett.},
  volume = {129},
  issue = {17},
  pages = {177601},
  numpages = {6},
  year = {2022},
  month = {Oct},
  publisher = {American Physical Society},
  doi = {10.1103/PhysRevLett.129.177601},
  url = {https://link.aps.org/doi/10.1103/PhysRevLett.129.177601}
}

@article{PhysRevB.108.085306,
  title = {{Four-dimensional topological Anderson insulator with an emergent second Chern number}},
  author = {Chen, Rui and Yi, Xiao-Xia and Zhou, Bin},
  journal = {Phys. Rev. B},
  volume = {108},
  issue = {8},
  pages = {085306},
  numpages = {8},
  year = {2023},
  month = {Aug},
  publisher = {American Physical Society},
  doi = {10.1103/PhysRevB.108.085306},
  url = {https://link.aps.org/doi/10.1103/PhysRevB.108.085306}
}

@article{PhysRevB.95.144406,
  title = {{Heat capacity evidence for proximity to the Kitaev quantum spin liquid in ${A}_{2} {\mathrm{IrO}}_{3}$ ($A=\mathrm{Na}$, Li)}},
  author = {Mehlawat, Kavita and Thamizhavel, A. and Singh, Yogesh},
  journal = {Phys. Rev. B},
  volume = {95},
  issue = {14},
  pages = {144406},
  numpages = {5},
  year = {2017},
  month = {Apr},
  publisher = {American Physical Society},
  doi = {10.1103/PhysRevB.95.144406},
  url = {https://link.aps.org/doi/10.1103/PhysRevB.95.144406}
}

@Article{Do2017,
author={Do, Seung-Hwan
and Park, Sang-Youn
and Yoshitake, Junki
and Nasu, Joji
and Motome, Yukitoshi
and Kwon, Yong Seung
and Adroja, D. T.
and Voneshen, D. J.
and Kim, Kyoo
and Jang, T.-H.
and Park, J.-H.
and Choi, Kwang-Yong
and Ji, Sungdae},
title={{Majorana fermions in the Kitaev quantum spin system $\alpha$-RuCl$_3$}},
journal={Nature Physics},
year={2017},
month={Nov},
day={01},
volume={13},
number={11},
pages={1079-1084},
issn={1745-2481},
doi={10.1038/nphys4264},
url={https://doi.org/10.1038/nphys4264}
}

@article{Winter2017,
doi = {10.1088/1361-648X/aa8cf5},
url = {https://dx.doi.org/10.1088/1361-648X/aa8cf5},
year = {2017},
month = {nov},
publisher = {IOP Publishing},
volume = {29},
number = {49},
pages = {493002},
author = {Stephen M Winter and Alexander A Tsirlin and Maria Daghofer and Jeroen van den Brink and Yogesh Singh and Philipp Gegenwart and Roser Valentí},
title = {{Models and materials for generalized Kitaev magnetism}},
journal = {Journal of Physics: Condensed Matter}
}

@Article{Wen2019,
author={Wen, Jinsheng
and Yu, Shun-Li
and Li, Shiyan
and Yu, Weiqiang
and Li, Jian-Xin},
title={{Experimental identification of quantum spin liquids}},
journal={npj Quantum Materials},
year={2019},
month={Apr},
day={05},
volume={4},
number={1},
pages={12},
issn={2397-4648},
doi={10.1038/s41535-019-0151-6},
url={https://doi.org/10.1038/s41535-019-0151-6}
}

@article{Panigrahi2024,
  title = {{Breakdown of order fractionalization in the CPT model}},
  author = {Panigrahi, Aaditya and Tsvelik, Alexei and Coleman, Piers},
  journal = {Phys. Rev. B},
  volume = {110},
  issue = {10},
  pages = {104520},
  numpages = {9},
  year = {2024},
  month = {Sep},
  publisher = {American Physical Society},
  doi = {10.1103/PhysRevB.110.104520},
  url = {https://link.aps.org/doi/10.1103/PhysRevB.110.104520}
}

@Article{Jin2023,
	title={{Lacing topological orders in two dimensions: Exactly solvable models for Kitaev's sixteen-fold way}},
	author={Jin-Tao Jin and Jian-Jian Miao and Yi Zhou},
	journal={SciPost Phys.},
	volume={14},
	pages={087},
	year={2023},
	publisher={SciPost},
	doi={10.21468/SciPostPhys.14.5.087},
	url={https://scipost.org/10.21468/SciPostPhys.14.5.087},
}

@Article{Jordan1928,
author={Jordan, P.
and Wigner, E.},
title={{{\"U}ber das Paulische {\"A}quivalenzverbot}},
journal={Zeitschrift f{\"u}r Physik},
year={1928},
month={Sep},
day={01},
volume={47},
number={9},
pages={631-651},
issn={0044-3328},
doi={10.1007/BF01331938},
url={https://doi.org/10.1007/BF01331938}
}

@article{Chen2008,
doi = {10.1088/1751-8113/41/7/075001},
url = {https://dx.doi.org/10.1088/1751-8113/41/7/075001},
year = {2008},
month = {feb},
publisher = {},
volume = {41},
number = {7},
pages = {075001},
author = {Han-Dong Chen and Zohar Nussinov},
title = {{Exact results of the Kitaev model on a hexagonal lattice: spin states, string and brane correlators, and anyonic excitations}},
journal = {Journal of Physics A: Mathematical and Theoretical}
}

@article{Yamada2021,
  title = {{Topological ${Z}_{2}$ invariant in Kitaev spin liquids: Classification of gapped spin liquids beyond projective symmetry group}},
  author = {Yamada, Masahiko G.},
  journal = {Phys. Rev. Res.},
  volume = {3},
  issue = {1},
  pages = {L012001},
  numpages = {7},
  year = {2021},
  month = {Jan},
  publisher = {American Physical Society},
  doi = {10.1103/PhysRevResearch.3.L012001},
  url = {https://link.aps.org/doi/10.1103/PhysRevResearch.3.L012001}
}

@article{S2015,
  title = {{Scattering Continuum and Possible Fractionalized Excitations in $\ensuremath{\alpha}\text{\ensuremath{-}}{\mathrm{RuCl}}_{3}$}},
  author = {Sandilands, Luke J. and Tian, Yao and Plumb, Kemp W. and Kim, Young-June and Burch, Kenneth S.},
  journal = {Phys. Rev. Lett.},
  volume = {114},
  issue = {14},
  pages = {147201},
  numpages = {5},
  year = {2015},
  month = {Apr},
  publisher = {American Physical Society},
  doi = {10.1103/PhysRevLett.114.147201},
  url = {https://link.aps.org/doi/10.1103/PhysRevLett.114.147201}
}

@Article{Glamazda2016,
author={Glamazda, A.
and Lemmens, P.
and Do, S.-H.
and Choi, Y. S.
and Choi, K.-Y.},
title={{Raman spectroscopic signature of fractionalized excitations in the harmonic-honeycomb iridates $\beta$- and $\gamma$-Li$_2$IrO$_3$}},
journal={Nature Communications},
year={2016},
month={Jul},
day={26},
volume={7},
number={1},
pages={12286},
issn={2041-1723},
doi={10.1038/ncomms12286},
url={https://doi.org/10.1038/ncomms12286}
}

@article{Banerjee2017,
author = {Arnab Banerjee  and Jiaqiang Yan  and Johannes Knolle  and Craig A. Bridges  and Matthew B. Stone  and Mark D. Lumsden  and David G. Mandrus  and David A. Tennant  and Roderich Moessner  and Stephen E. Nagler },
title = {{Neutron scattering in the proximate quantum spin liquid RuCl$_3$}},
journal = {Science},
volume = {356},
number = {6342},
pages = {1055-1059},
year = {2017},
doi = {10.1126/science.aah6015},
URL = {https://www.science.org/doi/abs/10.1126/science.aah6015}}

@article{PhysRevLett.114.077202,
  title = {{Hyperhoneycomb Iridate $\ensuremath{\beta}\text{\ensuremath{-}}{\mathrm{Li}}_{2}{\mathrm{IrO}}_{3}$ as a Platform for Kitaev Magnetism}},
  author = {Takayama, T. and Kato, A. and Dinnebier, R. and Nuss, J. and Kono, H. and Veiga, L. S. I. and Fabbris, G. and Haskel, D. and Takagi, H.},
  journal = {Phys. Rev. Lett.},
  volume = {114},
  issue = {7},
  pages = {077202},
  numpages = {5},
  year = {2015},
  month = {Feb},
  publisher = {American Physical Society},
  doi = {10.1103/PhysRevLett.114.077202},
  url = {https://link.aps.org/doi/10.1103/PhysRevLett.114.077202}
}

@article{M2009,
  title = {{Exactly solvable Kitaev model in three dimensions}},
  author = {Mandal, Saptarshi and Surendran, Naveen},
  journal = {Phys. Rev. B},
  volume = {79},
  issue = {2},
  pages = {024426},
  numpages = {8},
  year = {2009},
  month = {Jan},
  publisher = {American Physical Society},
  doi = {10.1103/PhysRevB.79.024426},
  url = {https://link.aps.org/doi/10.1103/PhysRevB.79.024426}
}

@Article{Lahtinen2017,
	title={{A Short Introduction to Topological Quantum Computation}},
	author={Ville Lahtinen and Jiannis K. Pachos},
	journal={SciPost Phys.},
	volume={3},
	pages={021},
	year={2017},
	publisher={SciPost},
	doi={10.21468/SciPostPhys.3.3.021},
	url={https://scipost.org/10.21468/SciPostPhys.3.3.021},
}

@article{Nayak2008,
  title = {{Non-Abelian anyons and topological quantum computation}},
  author = {Nayak, Chetan and Simon, Steven H. and Stern, Ady and Freedman, Michael and Das Sarma, Sankar},
  journal = {Rev. Mod. Phys.},
  volume = {80},
  issue = {3},
  pages = {1083--1159},
  numpages = {0},
  year = {2008},
  month = {Sep},
  publisher = {American Physical Society},
  doi = {10.1103/RevModPhys.80.1083},
  url = {https://link.aps.org/doi/10.1103/RevModPhys.80.1083}
}

@article{PhysRevLett.61.2015,
  title = {Model for a Quantum Hall Effect without Landau Levels: Condensed-Matter Realization of the "Parity Anomaly"},
  author = {Haldane, F. D. M.},
  journal = {Phys. Rev. Lett.},
  volume = {61},
  issue = {18},
  pages = {2015--2018},
  numpages = {0},
  year = {1988},
  month = {Oct},
  publisher = {American Physical Society},
  doi = {10.1103/PhysRevLett.61.2015},
  url = {https://link.aps.org/doi/10.1103/PhysRevLett.61.2015}
}

@article{PhysRevA.93.043827,
  title = {Synthetic dimensions in integrated photonics: From optical isolation to four-dimensional quantum Hall physics},
  author = {Ozawa, Tomoki and Price, Hannah M. and Goldman, Nathan and Zilberberg, Oded and Carusotto, Iacopo},
  journal = {Phys. Rev. A},
  volume = {93},
  issue = {4},
  pages = {043827},
  numpages = {17},
  year = {2016},
  month = {Apr},
  publisher = {American Physical Society},
  doi = {10.1103/PhysRevA.93.043827},
  url = {https://link.aps.org/doi/10.1103/PhysRevA.93.043827}
}

@article{PhysRevA.87.013814,
  title = {Four-dimensional photonic lattices and discrete tesseract solitons},
  author = {Juki\ifmmode \acute{c}\else \'{c}\fi{}, D. and Buljan, H.},
  journal = {Phys. Rev. A},
  volume = {87},
  issue = {1},
  pages = {013814},
  numpages = {4},
  year = {2013},
  month = {Jan},
  publisher = {American Physical Society},
  doi = {10.1103/PhysRevA.87.013814},
  url = {https://link.aps.org/doi/10.1103/PhysRevA.87.013814}
}

@article{PhysRevB.109.125303,
  title = {Four-dimensional Floquet topological insulator with an emergent second Chern number},
  author = {Liu, Zheng-Rong and Chen, Rui and Zhou, Bin},
  journal = {Phys. Rev. B},
  volume = {109},
  issue = {12},
  pages = {125303},
  numpages = {9},
  year = {2024},
  month = {Mar},
  publisher = {American Physical Society},
  doi = {10.1103/PhysRevB.109.125303},
  url = {https://link.aps.org/doi/10.1103/PhysRevB.109.125303}
}

@article{PhysRevLett.115.195303,
  title = {Four-Dimensional Quantum Hall Effect with Ultracold Atoms},
  author = {Price, H. M. and Zilberberg, O. and Ozawa, T. and Carusotto, I. and Goldman, N.},
  journal = {Phys. Rev. Lett.},
  volume = {115},
  issue = {19},
  pages = {195303},
  numpages = {6},
  year = {2015},
  month = {Nov},
  publisher = {American Physical Society},
  doi = {10.1103/PhysRevLett.115.195303},
  url = {https://link.aps.org/doi/10.1103/PhysRevLett.115.195303}
}

@article{
doi:10.1126/science.294.5543.823,
author = {Shou-Cheng Zhang  and Jiangping Hu },
title = {A Four-Dimensional Generalization of the Quantum Hall Effect},
journal = {Science},
volume = {294},
number = {5543},
pages = {823-828},
year = {2001},
doi = {10.1126/science.294.5543.823},
URL = {https://www.science.org/doi/abs/10.1126/science.294.5543.823},
eprint = {https://www.science.org/doi/pdf/10.1126/science.294.5543.823}}

@article{PhysRevB.94.041105,
  title = {Five-dimensional generalization of the topological Weyl semimetal},
  author = {Lian, Biao and Zhang, Shou-Cheng},
  journal = {Phys. Rev. B},
  volume = {94},
  issue = {4},
  pages = {041105},
  numpages = {5},
  year = {2016},
  month = {Jul},
  publisher = {American Physical Society},
  doi = {10.1103/PhysRevB.94.041105},
  url = {https://link.aps.org/doi/10.1103/PhysRevB.94.041105}
}

@article{PhysRevB.95.235106,
  title = {Weyl semimetal and topological phase transition in five dimensions},
  author = {Lian, Biao and Zhang, Shou-Cheng},
  journal = {Phys. Rev. B},
  volume = {95},
  issue = {23},
  pages = {235106},
  numpages = {6},
  year = {2017},
  month = {Jun},
  publisher = {American Physical Society},
  doi = {10.1103/PhysRevB.95.235106},
  url = {https://link.aps.org/doi/10.1103/PhysRevB.95.235106}
}

@article{yu_4d_2020-1,
	title = {{4D} spinless topological insulator in a periodic electric circuit},
	volume = {7},
	copyright = {http://creativecommons.org/licenses/by/4.0/},
	issn = {2095-5138, 2053-714X},
	url = {https://academic.oup.com/nsr/article/7/8/1288/5820240},
	doi = {10.1093/nsr/nwaa065},
	number = {8},
	urldate = {2025-09-08},
	journal = {National Science Review},
	author = {Yu, Rui and Zhao, Y X and Schnyder, Andreas P},
	month = aug,
	year = {2020},
	pages = {1288--1295},
}

@article{wang_circuit_2020,
	title = {Circuit implementation of a four-dimensional topological insulator},
	volume = {11},
	issn = {2041-1723},
	url = {https://www.nature.com/articles/s41467-020-15940-3},
	doi = {10.1038/s41467-020-15940-3},
	number = {1},
	urldate = {2025-09-08},
	journal = {Nature Communications},
	author = {Wang, You and Price, Hannah M. and Zhang, Baile and Chong, Y. D.},
	month = may,
	year = {2020},
	pages = {2356},
}

@article{PhysRevB.102.201111,
  title = {Microscopic models for Kitaev's sixteenfold way of anyon theories},
  author = {Chulliparambil, Sreejith and Seifert, Urban F. P. and Vojta, Matthias and Janssen, Lukas and Tu, Hong-Hao},
  journal = {Phys. Rev. B},
  volume = {102},
  issue = {20},
  pages = {201111},
  numpages = {5},
  year = {2020},
  month = {Nov},
  publisher = {American Physical Society},
  doi = {10.1103/PhysRevB.102.201111},
  url = {https://link.aps.org/doi/10.1103/PhysRevB.102.201111}
}

@article{PhysRevLett.107.087205,
  title = {Fermionic Magnons, Non-Abelian Spinons, and the Spin Quantum Hall Effect from an Exactly Solvable Spin-$1/2$ Kitaev Model with SU(2) Symmetry},
  author = {Yao, Hong and Lee, Dung-Hai},
  journal = {Phys. Rev. Lett.},
  volume = {107},
  issue = {8},
  pages = {087205},
  numpages = {5},
  year = {2011},
  month = {Aug},
  publisher = {American Physical Society},
  doi = {10.1103/PhysRevLett.107.087205},
  url = {https://link.aps.org/doi/10.1103/PhysRevLett.107.087205}
}

@article{PhysRevB.106.125144,
  title = {Order fractionalization in a Kitaev-Kondo model},
  author = {Tsvelik, Alexei M. and Coleman, Piers},
  journal = {Phys. Rev. B},
  volume = {106},
  issue = {12},
  pages = {125144},
  numpages = {16},
  year = {2022},
  month = {Sep},
  publisher = {American Physical Society},
  doi = {10.1103/PhysRevB.106.125144},
  url = {https://link.aps.org/doi/10.1103/PhysRevB.106.125144}
}

@article{wegnerDualityGeneralizedIsing1971,
  title = {Duality in {{Generalized Ising Models}} and {{Phase Transitions}} without {{Local Order Parameters}}},
  author = {Wegner, Franz J.},
  year = 1971,
  month = oct,
  journal = {Journal of Mathematical Physics},
  volume = {12},
  number = {10},
  pages = {2259--2272},
  issn = {0022-2488, 1089-7658},
  doi = {10.1063/1.1665530},
  urldate = {2026-02-02},
  language = {english}
}

@article{PhysRevB.92.115122,
  title = {Thermal fractionalization of quantum spins in a Kitaev model: Temperature-linear specific heat and coherent transport of Majorana fermions},
  author = {Nasu, Joji and Udagawa, Masafumi and Motome, Yukitoshi},
  journal = {Phys. Rev. B},
  volume = {92},
  issue = {11},
  pages = {115122},
  numpages = {6},
  year = {2015},
  month = {Sep},
  publisher = {American Physical Society},
  doi = {10.1103/PhysRevB.92.115122},
  url = {https://link.aps.org/doi/10.1103/PhysRevB.92.115122}
}

@article{PhysRevD.19.3682,
  title = {Phase diagrams of lattice gauge theories with Higgs fields},
  author = {Fradkin, Eduardo and Shenker, Stephen H.},
  journal = {Phys. Rev. D},
  volume = {19},
  issue = {12},
  pages = {3682--3697},
  numpages = {0},
  year = {1979},
  month = {Jun},
  publisher = {American Physical Society},
  doi = {10.1103/PhysRevD.19.3682},
  url = {https://link.aps.org/doi/10.1103/PhysRevD.19.3682}
}

@article{PhysRevB.92.115127,
  title = {Dynamics of fractionalization in quantum spin liquids},
  author = {Knolle, J. and Kovrizhin, D. L. and Chalker, J. T. and Moessner, R.},
  journal = {Phys. Rev. B},
  volume = {92},
  issue = {11},
  pages = {115127},
  numpages = {20},
  year = {2015},
  month = {Sep},
  publisher = {American Physical Society},
  doi = {10.1103/PhysRevB.92.115127},
  url = {https://link.aps.org/doi/10.1103/PhysRevB.92.115127}
}
 \end{document}